\documentclass[twoside]{ae100prg}
\usepackage{graphicx,color}
\usepackage{bm,graphics,graphicx,epsfig,color,ulem,amsfonts,amsmath}
\usepackage[breaklinks]{hyperref}
\usepackage{booktabs}

\def\be{\begin{equation}}
\def\ee{\end{equation}}
\def\ba{\begin{eqnarray}}
\def\ea{\end{eqnarray}}

\def\bse{\begin{subequations}}
\def\ese{\end{subequations}}

\begin{document}
\title{2.5PN kick from black-hole binaries in circular orbit: Nonspinning case}


\author{
Chandra Kant Mishra$^{1,2}$ \footnote[4]{Present address:
IISER Thiruvananthapuram, Computer Science
Building, College of Engineering Campus, 
Thiruvananthapuram, 695016, India}
, 
K G Arun$^3$, and 
Bala R Iyer$^{1}$
}

\address{$^1$ Raman Research Institute, Sadashivanagar, Bangalore, 560080, India}
\address{$^2$ Indian Institute of Science, Bangalore, 560012, India}
\address{$^3$ Chennai Mathematical Institute, Siruseri, Kelambakkam, 603103, India}

\email{cmishra@iisertvm.ac.in}
\email{kgarun@cmi.ac.in}
\email{bri@rri.res.in}

\begin{abstract}
Using the Multipolar post-Minskowskian formalism, we compute the linear momentum flux 
from black-hole binaries in circular orbits and having no spins. The 
total linear momentum flux contains various types of {\it instantaneous} (which 
are functions of the retarded time) and {\it hereditary} (which depends on the 
dynamics of the binary in the past) terms both of which are analytically 
computed. In addition to the inspiral contribution, we use a simple model of plunge 
to compute the kick or recoil accumulated during this phase.
\end{abstract}

\section{Introduction}
In classical radiation theory, any form of radiation that has a preferred direction (anisotropic emission) results in the {\it recoil} of the system in the opposite direction. Such a recoil has important consequences in astrophysics like the pulsar acceleration due to the radiation asymmetry~\cite{Harrison75}. This effect can be understood in terms of multipole moments of the system, more specifically interference between different multipoles~\cite{Peres62}. It was argued in Ref.~\cite{Peres62}, on general grounds, that the leading recoil effect in the case of electromagnetic radiation could come from the interference between electric dipole and electric quadrupole or magnetic dipole (See Ref.~\cite{HFHkick04} for a more detailed discussion). A similar effect can happen in the case of gravitational radiation~\cite{Peres62}. Since the leading emission in the case of Gravitational Waves (GWs) in GR is quadrupolar, the lowest order recoil effect in GR arises from the interference of mass quadrupole and mass octupole or current quadrupole.\footnote{Both mass octupole and current quadrupole have the same parity.} This GW recoil can be seen as a consequence of linear momentum flux emission by the radiating system.

Let us now focus on a particular type of GW source known as coalescing compact binaries (CCBs). As the name indicates, these are systems consisting of neutron stars (NS) or black holes (BH) which go around each other in a bound orbit. The orbit keeps on shrinking due to  gravitational radiation reaction until the two objects merge to form a compact object, most likely a BH. As is well known, the GW emission is not isotropic (due to the quadrupolar nature) and thus the emission is beamed in some direction. As the two bodies move in their respective orbits, this direction also keep changing. If the orbit is closed, then over an orbit, the binary's centre of mass would return to the same point where it started, making the net recoil zero. However, if the orbit is not closed, as is the case for CCBs, there can be an accumulation of recoil over these orbits until the two bodies coalesce into a single BH,
eventually manifesting as a kick to this newly formed BH. This is the GW recoil in the context of CCBs. 

Gravitational recoil has important implications for astrophysics, especially models of black hole formation and growth (see e.g.~\cite{Merritt04}). If the recoil velocity of a merging compact binary system is greater than the escape velocity of the galaxy, it can result in the ejection of the newly formed black hole from the galaxy. Hence very accurate estimation of the recoil velocity of the merging compact binaries is important to understand the astrophysics of black hole formation and growth.

Evolution of CCBs can be divided into three phases: adiabatic {\it inspiral}, nonlinear {\it merger} and perturbative {\it ringdown}. The inspiral part and ringdown phases can be accurately modelled analytically using post-Newtonian (PN) approximation to general relativity~\cite{Bliving} and BH perturbation theory~\cite{TSLivRev03}, respectively. Due to recent successes in numerical relativity (see Refs.~\cite{Pretorius07Review,CentrellaNRrev10} for  reviews), the highly nonlinear merger phase can be modelled quite accurately by numerically solving Einstein's equations.

In this work, we use the PN approximation to calculate the linear momentum loss and resulting GW recoil from a  black-hole binary\footnote{Note that the expressions for the linear momentum flux (Eq.~\eqref{eq-circlmf:lmf}) and that for the recoil velocity (Eq.~\eqref{eq-circlmf:recvel}) are even applicable to compact binary systems involving NSs as components. However, the plunge computations presented here assume that both the components of the binary are BHs.} moving in circular orbits up to 2.5PN order extending the earlier works of Fittchett (Newtonian order)~\cite{Fitchett83}, Wiseman (1PN order~\cite{Wi92}) and Blanchet et al. ~(through 2PN~\cite{BQW05}). A more detailed account of the contents of this paper can be found in Ref.~\cite{MAI12}.

This article is organized in the following way. In Sec.~\ref{MPM}, we write down the multipolar expansion of the linear momentum flux and the final expression for the flux in terms of source multipole moments. Section~\ref{Recoil} derives the 2.5PN expression for the recoil and Sec.~\ref{NumEst} discusses the numerical estimates of recoil. Conclusions are given in Sec.~\ref{Conclusion}.
\section{Multipole Expansion for Linear momentum flux}\label{MPM}
The general formula for linear momentum flux in the far-zone of the source
in terms of symmetric trace-free (STF) radiative multipole moments
is given in \cite{Th80} and at relative 2.5PN order it takes the following form (see e.g. Eq.~(4.20') of Ref.~\cite{Th80}.) 
\begin{eqnarray}
\label{eq:structure-LMF}
{\mathcal{F}_{P}^{i}}(U)&=& \frac{G}{c^7}\,\biggl\{\left[\frac{2}{63}\,
U^{(1)}_{ijk}\,U^{(1)}_{jk}+\frac{16}{45}\,\varepsilon_{ijk}
U^{(1)}_{ja}\,V^{(1)}_{ka}\right]\nonumber\\&&
+{1\over c^2}\left[\frac{1}{1134}\,U^{(1)}_{ijkl}\,U^{(1)}_{jkl}
+\frac{1}{126}\,\varepsilon_{ijk}
U^{(1)}_{jab}\,V^{(1)}_{kab}
+\frac{4}{63}\,V^{(1)}_{ijk}\,V^{(1)}_{jk}\right]\nonumber\\&&
+{1\over
c^4}\left[\frac{1}{59400}\,U^{(1)}_{ijklm}\,U^{(1)}_{jklm}+\frac{2}{14175}\,\varepsilon_{ijk}
U^{(1)}_{jabc}\,V^{(1)}_{kabc}\right.\nonumber\\&&\left.+
\frac{2}{945}\,V^{(1)}_{ijkl}\,V^{(1)}_{jkl}\right]
+{\cal O}\left({1\over c^6}\right) \biggr\}.
\end{eqnarray}

In the equation above $U_L$ and $V_L$ denote the mass and current type radiative multipoles of the source (suffix $L$ captures the multi-index structure of the $U$ and $V$ moments) and $U_L^{(n)}$ and $V_L^{(n)}$ denote the $n^{\rm th}$ time derivatives of the moments. ${\cal O}(1/c^6)$ denotes the omission of terms higher than 3PN w.r.t the leading term. The moments which appear in the above formula are functions of retarded time $T-\frac{R}{c}$, where $R$ denotes the distance of the source relative to the observer and $T$ the time of observation, both in radiative coordinates. In the 
MPM formalism $U_{L}$ and $V_{L}$ are related to canonical moments $M_L$ and 
$S_L$ (Eqs. (5.4)-(5.8) of \cite{BFIS08}) which in turn are related to source 
moments $\left\{I_L, J_L, X_L, W_L, Y_L, Z_L\right\}$ (Eqs. (5.9)-(5.11) of \cite{BFIS08}). Using these inputs, one can re-express the radiative multipole moments in terms of source moments. Then one notices that the total linear momentum flux consists of two parts: one type of terms are functions of the retarded time called {\it instantaneous terms} and the other which are sensitive to the dynamics of the source in its entire past called as {\it hereditary terms}. Explicit expressions for the two types of terms is given in Eqs.~(2.3)-(2.5) of Ref.~\cite{MAI12}. Further, the explicit expressions for various source multipole moments are given in Eqs.~(3.1)-(3.4) of Ref.~\cite{MAI12}. The only other input we require is the 2.5PN accurate equations of motion which can be found in Ref.~\cite{BFIS08,K08}. Using these inputs, and working in {\it harmonic coordinates}, 
we obtain the total linear momentum flux in terms of gauge independent variable $x=(m\omega)^{2/3}$ as
\begin{eqnarray}
\label{eq-circlmf:lmf}
{{\mathcal F}_{P}^i}&=&
-{464\over105}\,{c^4\over G}\,\sqrt{1-4\,\nu}\,x^{11/2}\,\nu^2
\left\{\left[
1+x\left(-{452\over87}-{1139\over522}\,\nu\right)
\right.\right.\nonumber\\&&\left.\left.
+{309\over58}\,\pi\,x^{3/2}
+x^2\left(-{71345\over22968}+{36761\over2088}\,\nu
+{147101\over68904}\,\nu^2\right)
\right.\right.\nonumber\\&&\left.\left.
+x^{5/2}\left(-{2663\over116}\,\pi-{2185\over87}\,\pi\,\nu\right)
\right]\,{\hat{\lambda}^{i}}
+x^{5/2}\left[-\frac{106187}{50460}
\right.\right.\nonumber\\
&&\left.\left.
+\frac{32835}{841}\,{\log\,2}
-\frac{77625}{3364}\,{\log\,3}
+\left(\frac{10126}{4205}
-\frac{109740}{841}\,{\log\,2}
\right.\right.\right.\nonumber\\&&\left.\left.\left.
+\frac{66645}{841}\,{\log\,3}\right)\,\nu
\right]{\hat{n}_i}+{\cal O}\left({1\over c^6}\right)
\right\}.
\end{eqnarray}
In the equation above ${\bf \hat{n}}$ and $\hat{\bm \lambda}$ are related to the phase angle $\psi$ as 
\begin{eqnarray}
\label{eq:nlpsi}
{\bf \hat{n}}&={\cos{\psi}}\,{\bf \hat{e}_x}+{\sin{\psi}}\,{\bf \hat{e}_y}\,,
\label{eq:n-psi}\\
\hat{\bm \lambda}&=-{\sin{\psi}}\,{\bf \hat{e}_x}+{\cos{\psi}}\,{\bf \hat{e}_y}\,,
\label{eq:l-psi}
\end{eqnarray}
\section{Computation of the Recoil}\label{Recoil}
Having computed the linear momentum flux for compact binaries in circular
orbits, one can obtain the recoil velocity by integrating the momentum balance equation
\begin{equation}
{dP^i\over dt}=-{\mathcal F}_P^i\,.
\end{equation}
to get
\begin{equation}
\label{eq:deltamom}
{\Delta P^i}=-\int_{-\infty}^{t}\,dt'\,{\mathcal F_P^i}\,.
\end{equation}
Performing the integration we find,
{\begin{eqnarray}
\label{eq-circlmf:recvel}
V_{\rm recoil}^{i}&=&{464\over105}\,c\,\sqrt{1-4\,\nu}\,x^4\,\nu^2\biggl\{\left[1+x\left(-{452\over87}-{1139\over522}\,\nu\right)
+{309\over58}\,\pi\,x^{3/2}\right.\nonumber\\&&\left.
+x^2\left(-{71345\over22968}+{36761\over2088}\,\nu+{147101\over68904}\,\nu^2\right)
+x^{5/2}\left(-{2663\over116}\,\pi
\right.\right.\nonumber\\&&\left.\left.
-{2185\over87}\,\pi\,\nu\right)\right]\,{\hat{n}_i}
+x^{5/2}\left[\frac{106187}{50460}-\frac{32835}{841}\,{\log\,2}
+
\frac{77625}{3364}\,{\log 3}
\right.\nonumber\\&&\left.
+\left(\frac{41034}{841}
+\frac{109740}{841}\,{\log\,2}-\frac{66645}{841}\,{\log\,3}
\right)\nu\right]{\hat{\lambda}_i}\nonumber\\&&
+{\cal O}\left({1\over c^6}\right)
\biggr\}.
\end{eqnarray}}
\section{Numerical estimates of the recoil velocity including plunge contribution}\label{NumEst}
As is evident from Eq.~(\ref{eq-circlmf:recvel}), the recoil velocity depends on the (symmetric) mass ratio $\nu$ and does {\it not} depend on the total mass.
This is consistent with our understanding that the origin of the recoil is in the mass asymmetry. We finally want to numerically estimate the recoil velocities as a function of $\nu$. Our calculation based on 2.5PN approximation yields a maximum kick velocity of $\sim 4$km/sec as opposed to $\sim 22$ km/sec of the 2PN model given in Ref.~\cite{BQW05}. Thus 2.5PN estimates predict a smaller kick velocity than the 2PN model. This result is obtained by integrating the linear momentum flux till the Innermost Stable Circular Orbit (ISCO), up till which PN approximation is considered to be valid.

 Since the dominant contribution to the recoil comes towards the late stages of the binary evolution, we incorporate the contribution from the `plunge' phase of the evolution beyond the last stable orbit, following and extending a model proposed in Ref.~\cite{BQW05}. The method may be considered to be less sophisticated version of the Effective One Body approach~\cite{BuonD98,BuonD00}. In this model, the plunge can  be viewed as that of a test particle moving in the fixed Schwarzshild geometry of mass $m$. The contribution from the plunge phase is estimated using the PN formulae assuming they are valid even beyond ISCO. Since the PN representation is usually not reliable inside ISCO, this should be a source of error and in general this computation is only a crude estimate.
\begin{figure}[t]
\begin{center}
\includegraphics[scale=0.3]{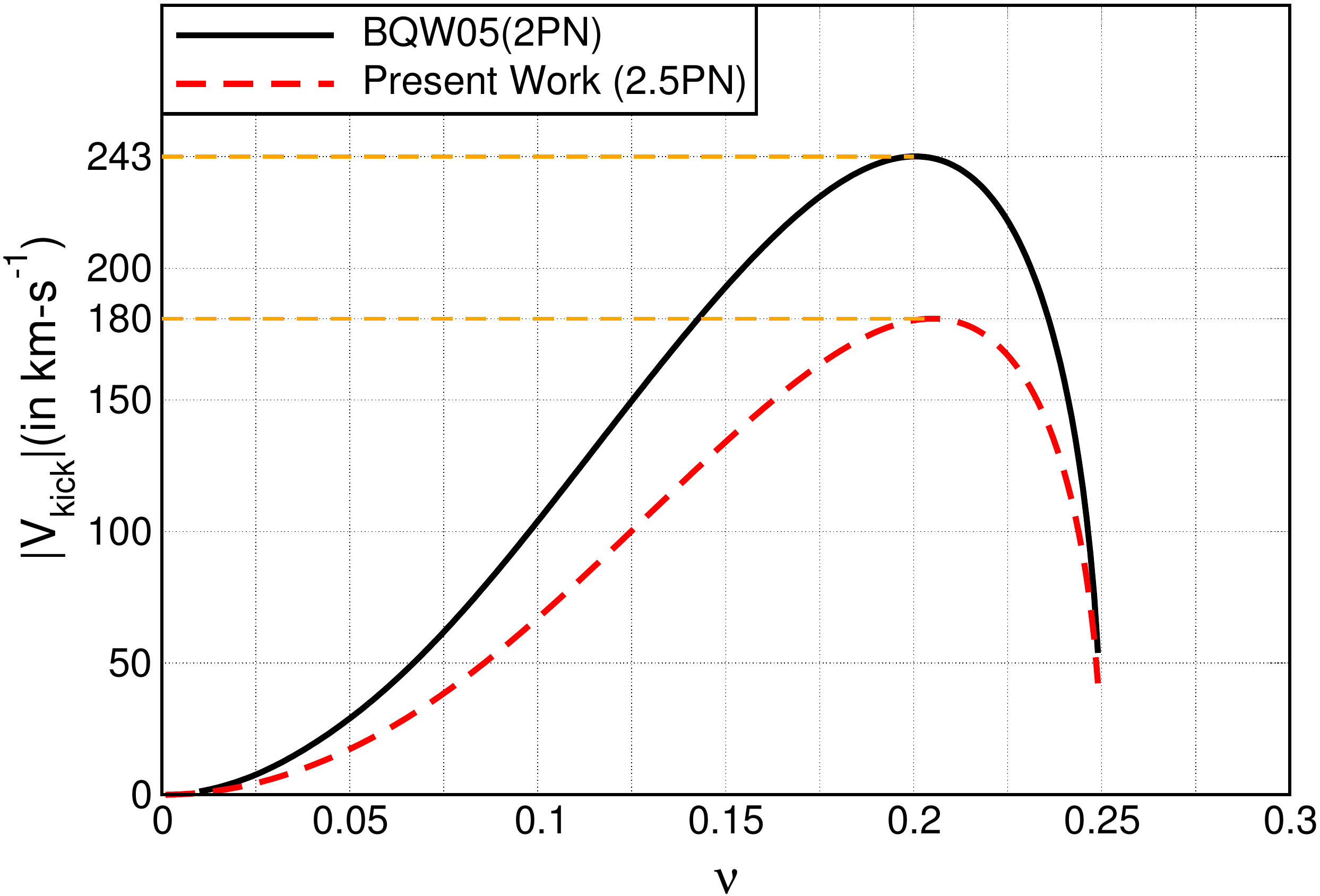}
\end{center}
\caption{Kick velocity imparted to the remnant of the compact binary coalescence as a function of symmetric mass ratio. This figure incorporates the contribution from the plunge phase of the binary evolution. The results of 2.5PN+plunge is compared against that of Ref.~\cite{BQW05} where the estimation was done with a 2PN+plunge model.}
\label{RecoilPlot}
\end{figure}
Further, the the recoil velocity accumulated during the two phases (inspiral and the plunge) can be obtained by taking a vector sum of the two estimates. This is achieved by matching the circular orbit at the ISCO to a suitable plunge orbit. The final results of the numerical estimates of the recoil velocity is presented in Fig~\ref{RecoilPlot}. We compare the recoil velocity based on our 2.5PN inspiral + plunge model with that of 2PN inspiral + plunge model of Ref.~\cite{BQW05}. (Specifications are same as those of Fig.~1 of \cite{BQW05}.) As is evident, our 2.5PN inspiral + plunge model predicts smaller recoil velocity for almost all values of $\nu$. The maximum of the curve drops from $\sim 243$ km/s of \cite{BQW05} to $\sim 180$ km/s. This may be attributed to the oscillatory convergence of the PN series observed in various contexts (see e.g.~\cite{DIS01}).
While our estimates are within the error bars given by \cite{BQW05}, one should keep in mind that there are also error bars associated with our estimates due to systematic effects such as neglect of higher order PN contributions. 

\section{Conclusion}\label{Conclusion}
Using the MPM formalism, we computed the linear momentum flux due to GW emission
from inspiralling compact binaries moving in circular orbit up to 2.5PN order.
This complements the computations of energy flux ~\cite{B96}, waveform and polarizations~\cite{ABIQ04,KBI07} at 2.5PN order for circular orbits. Using the PN linear momentum flux and an analytical model of plunge~\cite{BQW05}, we have estimated the contributions to the recoil from the inspiral and plunge phase, as a function of symmetric mass ratio.
\section*{References}
\bibliography{/home/arun/tphome/arun/LMF-Circular/ref-list}
\end{document}